\documentclass[11pt]{article}
\usepackage{graphicx}

\newcommand{\Kvec}{$K^*(892)$ \hspace{-0.07cm}}
\newcommand{\Kscal}{$K^*_0(1430)$ \hspace{-0.07cm}}

\title{\bf{Final state interactions and CP violation in $B$ decays to three pseudoscalars}
\thanks{This work has been supported in part by the Polish Ministry of Science
and Higher Education (grant No N N202 248135) and by the IN2P3-Polish
Laboratories Convention (project No 08-127).}}

\author{R. Kami\'nski$^a$, B. El Bennich$^b$, A.~Furman$^c$, L.~Le\'sniak$^a$, B.~Loiseau$^d$,
B.~Moussallam$^e$\\ \\
\textit{
$^a$Henryk Niewodnicza\'nski Institute of Nuclear Physics, Polish Academy of Sciences,}\\
\textit{
 ul. Radzikowskiego 152, 31-342 Krak\'ow, Poland,}\\
\textit{
$^b$Physics Division, Argonne National Laboratory, Argonne, Illinois,
 60439, USA,}\\
\textit{
$^c$ul. Bronowicka 85/26, 30-091 Krak\'ow, Poland,}\\
\textit{
$^d$Laboratoire de Physique Nucl\'eaire et de Hautes \'Energies}\\ 
\textit{
(IN2P3--CNRS--Universit\'es Paris 6 et 7),}\\ 
\textit{
Groupe Th\'eorie,
Universit\'e Pierre et Marie Curie, 4 place Jussieu, 75252 Paris, France,}\\
\textit{
$^e$Groupe de Physique Th\'eorique, Institut de Physique Nucl\'eaire 
(IN2P3--CNRS),}\\
\textit{
Universit\'e Paris-Sud 11, 91406 Orsay Cedex, France}\\}

\date{}

\begin{document}

\maketitle

\begin{abstract}
We study CP violation and final state interactions between pions and kaons
in $B^+, \,B^-, -\,B^0$ and $\bar B ^0$ decays into $K\pi\pi$. 
The weak transition amplitudes consist of two terms: the first part is
derived in QCD factorization approach and the second one is a phenomenological long-distance
charming penguin contribution.
The final state $K \pi$ interactions in $S$- and $P$-waves are 
described by strange scalar and vector form factors, respectively.
These are determined using 
a unitary coupled channel model together with chiral symmetry and asymptotic QCD constraints.
The final state interactions are dominated by presence of the scalar 
$K^*_0(1430)$ and the vector $K^{*}(892)$ resonances.
We show that additional charming penguin amplitudes are needed to reproduce the latest experimental 
$K \pi$ effective mass and helicity angle distributions,
branching fractions and asymmetries obtained by Belle and BaBar collaborations.
\end{abstract}


\section{Introduction}
Recent detailed Dalitz plot analyzes for different $B\to K\pi^+\pi^- $ 
decays were performed by BaBar and Belle collaborations
\cite{Garmash:2005rv,Garmash2007,Aubert:2008bj,Aubert:2007bs,Abe2005}.
New and rich data on $K\pi$ effective mass ($m_{K\pi}$) and helicity angle distributions, 
branching fractions and CP asymmetries have been delivered.
The effective mass distributions show distinct accumulation of events below 2 GeV around the scalar 
$K^*_0(1430)$ and the vector  $K^*(892)$ resonances.
In experimental analyzes the event distributions have been studied using the isobar model with
decay amplitudes parameterized by sums of Breit-Wigner terms and a background.
In \cite{Aubert:2008bj,Aubert:2007bs} 
an effective range component has been introduced in the $K\pi$ $S$-wave amplitude.

In this short report we present main results of our work on $B$ decays into three mesons, 
$B\to K\pi^+\pi^-$, with final state interactions between  kaons and pions in the $S$- and $P$-waves \cite{BFKLLM}.
In our analysis we restrict ourselves  to the low effective $K\pi$ mass region $m_{K\pi} < 2$ GeV
where most of the $K\pi$ resonant structures are seen.
It is reasonable to expect that in such a kinematical configuration 
three-body interactions are suppressed.
This allows us to grant  the validity of quasi two-body factorization approach.
We also assume that the $K \pi$ pairs originate from a quark-antiquark state.
Main assumption in our approach is, however, that
the corrections to factorization can be absorbed into 
$m_{K\pi}$ independent modifications of the QCD Wilson coefficients. 
We take parts of these corrections from quasi-two-body 
calculations and append a phenomenological part.   

In the factorization approach the decay amplitudes can be expressed as a product 
of the effective QCD part and  the two matrix elements of the vector currents 
which involve two functions of $m_{K\pi}^2$: the strange scalar and the vector form factors. 
The introduction of form factors, constrained by  theory and experiments other than $B$ decays, 
is an alternative to the isobar model often used in experimental analyzes.
This latter approximation violates unitarity what can, for example, lead to a distortion of 
information on resonances present in the final states.

\section{Theoretical model}

The weak amplitudes for the $B^{\pm}$, $B^0$ and $\bar B^0$ decays, have one term derived in the QCD factorization, 
and the second one being a phenomenological contribution to the penguin amplitudes.
For example, the $K \pi$ $S$-wave contribution to the $B^- \to K^- \pi^+\pi^-$ decay amplitude reads
(expressions for other $B$ decay amplitudes can be found in \cite{BFKLLM}):
\begin{eqnarray}
\nonumber 
\hspace{-9.cm}\mathcal{A}_S^-(m_{K\pi}^2) = 
 \frac{G_F}{\sqrt{2}} (M_B^2-m_{\pi}^2)\frac{m_{K}^2-m_{\pi}^2}{m_{K\pi}^2}
f_0^{B^-\pi^-}(m_{K\pi}^2) \ {f_0^{K^-\pi^+}(m_{K\pi}^2)} \\ \nonumber 
\times 
 \bigg\{
\lambda_u\left(a_4^{u}(S)-\frac{a_{10}^{u}(S)}{2}+{ c_4^{u}}\right)
+\lambda_c\left(a_4^c(S)-\frac{a_{10}^c(S)}{2}+{ c_4^{c}}\right) \\ 
- \frac{2m_{K\pi}^2}{(m_b-m_d)(m_s-m_d)}
\times \left[
\lambda_u\left(a_6^{u}(S)-\frac{a_8^{u}(S)}{2}+{ c_6^{u}}\right)
+\lambda_c\left(a_6^c(S)-\frac{a_8^c(S)}{2}+{ c_6^{c}}\right)
\right]\bigg\}, 
\end{eqnarray}
where the $a_i$ ($i=4,6,8,10$) are the leading order factorization coefficients with 
$O(\alpha_s)$ vertex and penguin corrections
added in the quasi-two-body approximation of  pseudoscalar-scalar or pseudoscalar-vector final states.
The $c_i$ coefficients are additional phenomenological complex parameters fitted to the experimental data.
They can be interpreted as contributions from 
charming penguins, hard spectator interaction in the perturbative regime and as annihilation terms. 
$G_F$ denotes the Fermi coupling constant, $M_B,\ m_K$ and $m_\pi$ are the
masses of the charged $B$ mesons, kaons and pions, $f_0^{B^-\pi^-}(m_{K\pi}^2)$ and $f_0^{K^-\pi^+}(m_{K\pi}^2)$
are the $B^-$ to $\pi^-$ and the $K\pi$ scalar form factors respectively and $m_b,\ m_s$ and $\ m_d$ are $b,~ s$ and
$d$ quark masses.
The $\lambda_u= V_{ub} V^*_{us}$ and  $\lambda_c= V_{cb} V^*_{cs}$ are products of the 
Cabibbo-Kobayashi-Maskawa quark-mixing matrix elements $V_{qq'}$.
As we have shown in \cite{BFKLLM}, the $m_{K\pi}^2$ dependence of the $S$-wave amplitude
(3rd line of the Eq. (1)) plays an important role in 
the description of the sizable enhancement in the $K\pi$ mass distributions below 1 GeV. 
In the isobar model the corresponding $m_{K\pi}$ dependence is approximated by one fitted constant parameter.
Such approximation may not be very good for the wide \Kscal resonance since the value of $m_{K\pi}^2$ varies by a factor almost
equal to 10 from the $K\pi$ threshold to the limit of our calculations about 1.8 GeV.

The scalar and vector $K\pi$ form factors are connected to the scattering amplitudes 
in the $S$- and $P$-waves via unitarity relations. In the $S$-wave 
we treat two coupled $ K\pi$ and $K\eta '$ channels and in the $P$-wave three ones:
$ K\pi$, $K^*\pi$ and $K\rho$. 
The corresponding scattering amplitudes are constrained by experimental data,
especially by the LASS results obtained at SLAC, and by theoretical constraints at low effective mass. 
The Muskhelishvili-Omnes equations are used to calculate the $K \pi$ form factors. 


\section{Numerical results}

The $ K\pi$ effective mass and helicity angle distributions, 
branching fractions, CP asymmetries and the phase difference between the $B^0$ and
$\bar B^0$ decay amplitudes to $K^*(892)\pi$ are calculated from our model and fitted,
using the minimization program with four phenomenological free parameters $c_{4,6}^{u,c}$, 
to 319 data obtained by BaBar and Belle collaborations.
Branching fractions play a special role as they determine the absolute size of the decay amplitudes.
As an input 
we use the branching fractions for the ${B} \to K^*(892)\pi$ decays which are well determined since 
the $K\pi$ $P$-wave amplitude below 1 GeV is entirely dominated by the 
$K^*(892)$ resonance.
Contrary to the $P$-wave amplitude, the $S$-wave amplitude contains not only the resonant part
related to the wide $K^*_0(1430)$ but also a very broad $K^*_0(800)$ below 1 GeV.
This complexity of the $S$-wave results  in different parameterizations of the amplitudes 
in the BaBar and Belle analyzes.
Therefore we do not use the experimental branching fractions for the 
${B} \to K^*_0(1430)\pi$, which are not well determined, being largely model dependent.
Some of our model predictions for branching fractions in both waves are given in Table~1. 
These numbers have been calculated by integration of the double differential branching fractions 
in limited ranges of $m_{K\pi}$ around the \Kscal and \Kvec resonances.
The model branching fractions for the $B \to K^*(892) \pi$ decays agree   
quite well  with  the corresponding experimental values within their errors.
If the phenomenological parameters $c_i^p\ (i=4,6$ and $p=u,c$) 
are equal to zero, than 
the theoretical branching fractions for the $P$-wave are 4-5 times smaller than the experimental ones
and those for the $S$-wave differ by a factor of 2. 
One can notice that the theoretical results for the $S$-wave 
are lower than the experimental data, being substantially smaller than the 
Belle results but closer to the BaBar numbers. 
\begin{table}[h!]
\begin{center}
\caption{Branching fractions averaged over charge conjugate $B\to K\pi\pi$ decays 
in units of $10^{-6}$. $(K^+\pi^-)_P$ and $(K^+\pi^-)_S$ denote the
$(K^+\pi^-)$ pair in $P$-wave and $S$-wave, respectively. 
The values of the model, calculated by the integration over the given 
$m_{K\pi}$ range,
 are compared to the corresponding Belle ~\cite{Garmash:2005rv,Garmash2007} and 
BaBar ~\cite{Aubert:2008bj,Aubert:2007bs} results.
 The model errors are the phenomenological parameter uncertainties 
 found in the minimization procedure.
}
\begin{tabular}[]{|c|c|c|c|c|}
\hline
decay mode   & $m_{K\pi}$ range (GeV) & Belle          &  BaBar       &  our model\\
\hline
 $B^+\to (K^+\pi^-)_P\ \pi^-$ &$(0.82,0.97)$   & $5.35\pm0.59$  & $5.98\pm0.75$ &
 $5.73\pm0.14$ \\
\hline
 $B^0\to (K^0\pi^+)_P\ \pi^-$ &$(0.82,0.97)$   & $4.65\pm0.77$  & $6.47\pm0.75$ &
 $5.42\pm0.16$ \\
\hline
 $B^+\to (K^+\pi^-)_S\ \pi^-$ &$(0.64,1.76)$   & $27.0\pm2.5$  & $22.5\pm4.6$ &
 $16.5\pm0.8$ \\
\hline
 $B^0\to (K^0\pi^+)_S\ \pi^-$ &$(0.64,1.76)$  & $26.0\pm3.4$  & $17.3\pm4.6$ &
 $15.8\pm0.7$ \\
\hline
\end{tabular}
\end{center}
\end{table}

In Table~2 we present comparison of our results for the CP asymmetries with experimental ones.
For $B^- \to (K^-\pi^+)_P\pi^-$  decays our asymmetries lie between those of Belle and BaBar. 
The results for $B^-\to (K^-\pi^+)_S\ \pi^-$  and for $\bar B^0\to (\bar K^0\pi^-)_S\ \pi^+$ 
decays agree with the experimental values of both collaborations.

\begin{table}[h!]
\begin{center}
\caption{Direct $CP$ asymmetries averaged over charge conjugate reactions. 
The values of the model, calculated over the indicated $m_{K\pi}$ range,  
are compared to the Belle \cite{Garmash:2005rv} and BaBar \cite{Aubert:2008bj,Aubert:2007bs} results.
 The model errors are the phenomenological parameter uncertainties 
 found in the minimization procedure.
}
\begin{tabular}{|c|c|c|c|c|}
\hline
Decay mode & $m_{K\pi}$ range (GeV) & Belle & BaBar & our model \\
\hline
$B^-\to (K^-\pi^+)_P \pi^-$ & $(0.82,0.97)$ &  $-14.9\pm 6.8$ & $3.2\pm 5.4$  & $-2.5\pm 1.3$ \\ 
\hline
$\bar B^0\to (\bar K^0\pi^-)_P\ \pi^+$ & $(0.82,0.97)$ & $----$ & $14 \pm 12$ & $-19.6 \pm 3.0$\\ 
\hline
$B^-\to (K^-\pi^+)_S \pi^-$ & $(1.0,1.76)$  &$7.6\pm 4.6$ & $3.2\pm 4.6$  & $5.4 \pm 1.0$ \\ 
\hline
$\bar B^0\to (\bar K^0\pi^-)_S \pi^+$ & $(1.0,1.76)$ & $----$ & $17\pm 26$ & $-0.2 \pm 1.3$ \\
\hline
\end{tabular}
\end{center}
\end{table}

\begin{figure}[h!]
\begin{center}
  \includegraphics[angle=0,scale=0.44]{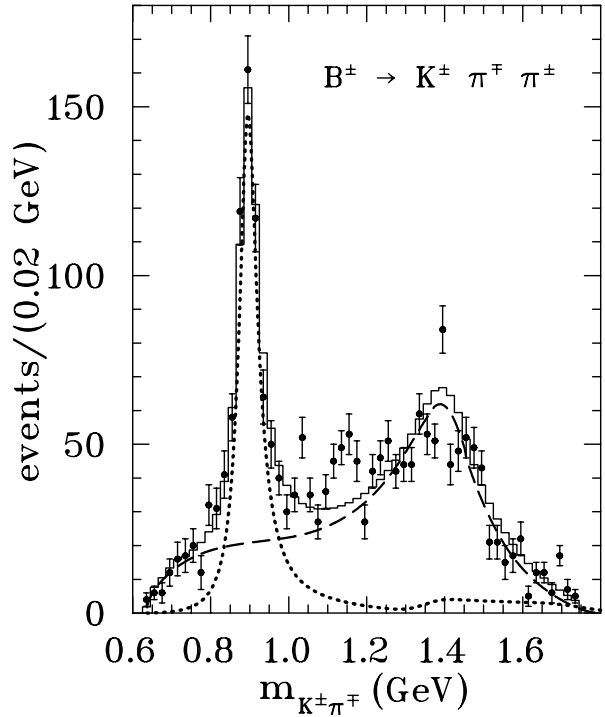}
  \caption{The  $K^\pm \pi^\mp$ effective mass distributions in the  
  $B^\pm \to K^\pm \pi^\mp \pi^\pm$ decays from the fit to the experimental data \cite{Garmash:2005rv}. 
  The
dashed line corresponds to the $S$-wave contribution of our model, 
the dotted line that of the $P$-wave and  the  histogram represents the coherent sum of the $S$- and $P$-wave contributions.
} 
\label{Fig.MassDistr}
\end{center}
\end{figure}

\begin{figure}[h!]
\begin{center}
  \includegraphics[angle=0,scale=0.40]{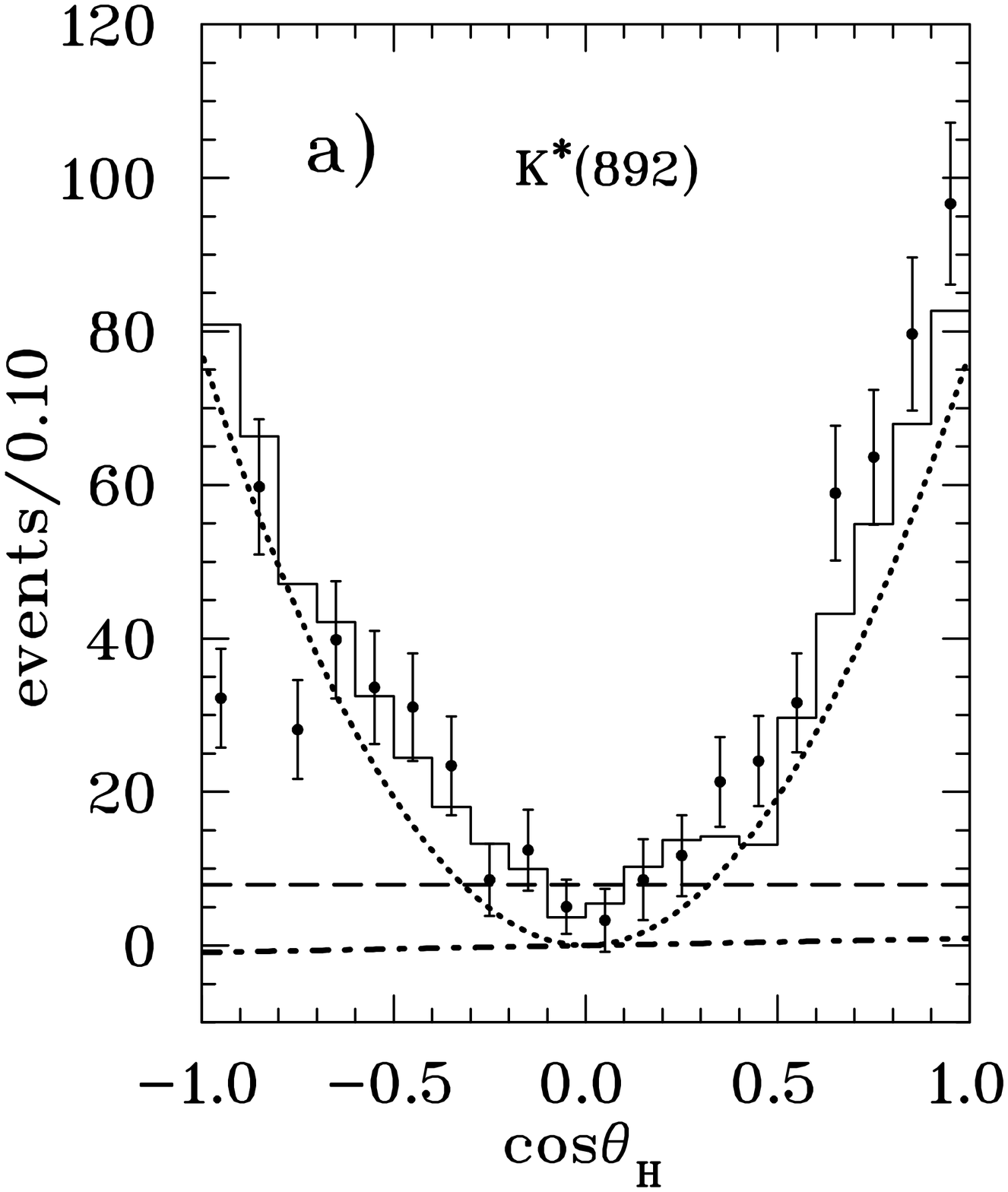}
  \includegraphics[angle=0,scale=0.40]{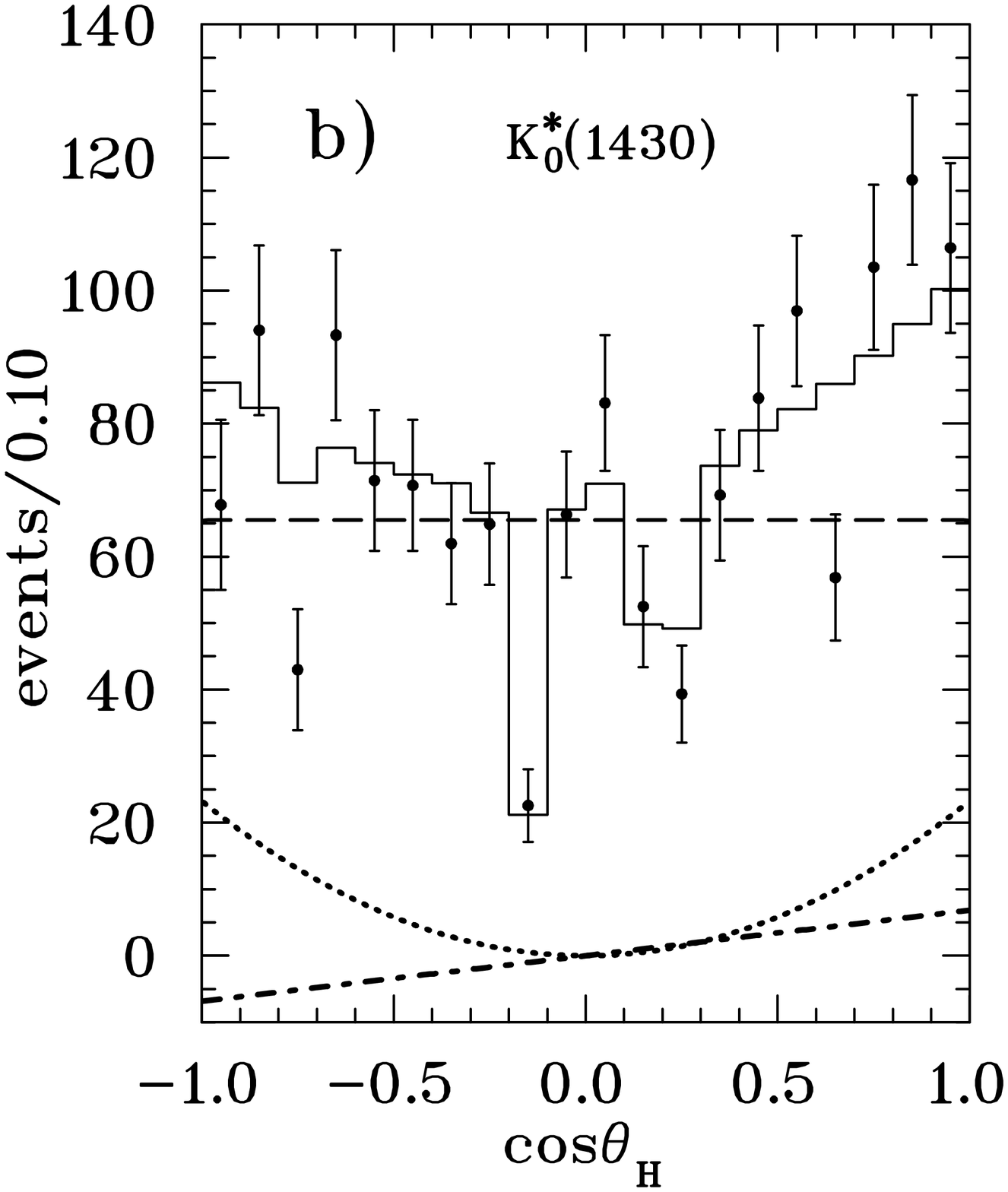}\\  
  \caption{Helicity angle distributions for  $B^\pm \to K^\pm \pi^\mp \pi^\pm$ decays  calculated
  from the averaged double differential distribution  integrated over $m_{K^\pm \pi^\mp}$ mass from
  0.82 to 0.97 GeV in the \Kvec case   a) and  from 1.0 to 1.76 GeV in the \Kscal case b).  
 Data  points are from  Ref.~\cite{Abe2005}. 
Dashed lines represent the $S$-wave contribution of our model, dotted lines that of the $P$-wave and  
the dot-dashed lines that of the interference term.
The histograms correspond to the sum of these three contributions.
} 
\label{Fig.Helicity}
\end{center}
\end{figure}

The fit on the model parameters leads to a good agreement with the experimental data, particularly 
for the kaon-pion effective mass and helicity angle distributions. 
Examples of these results are presented in 
Figs \ref{Fig.MassDistr} and \ref{Fig.Helicity}.

\begin{figure}[h!]
\begin{center}
  \includegraphics[angle=0,scale=0.43]{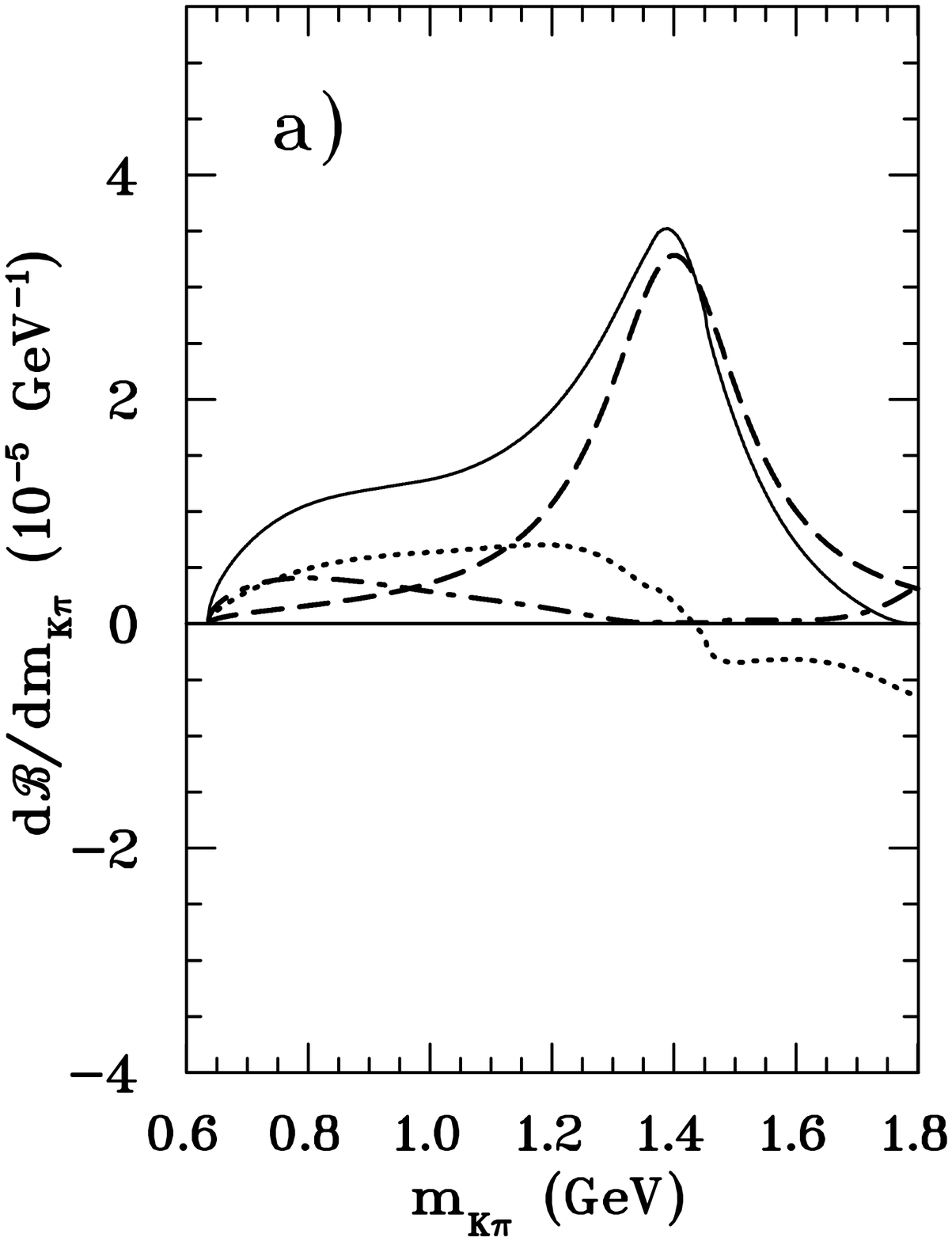}
  \includegraphics[angle=0,scale=0.43]{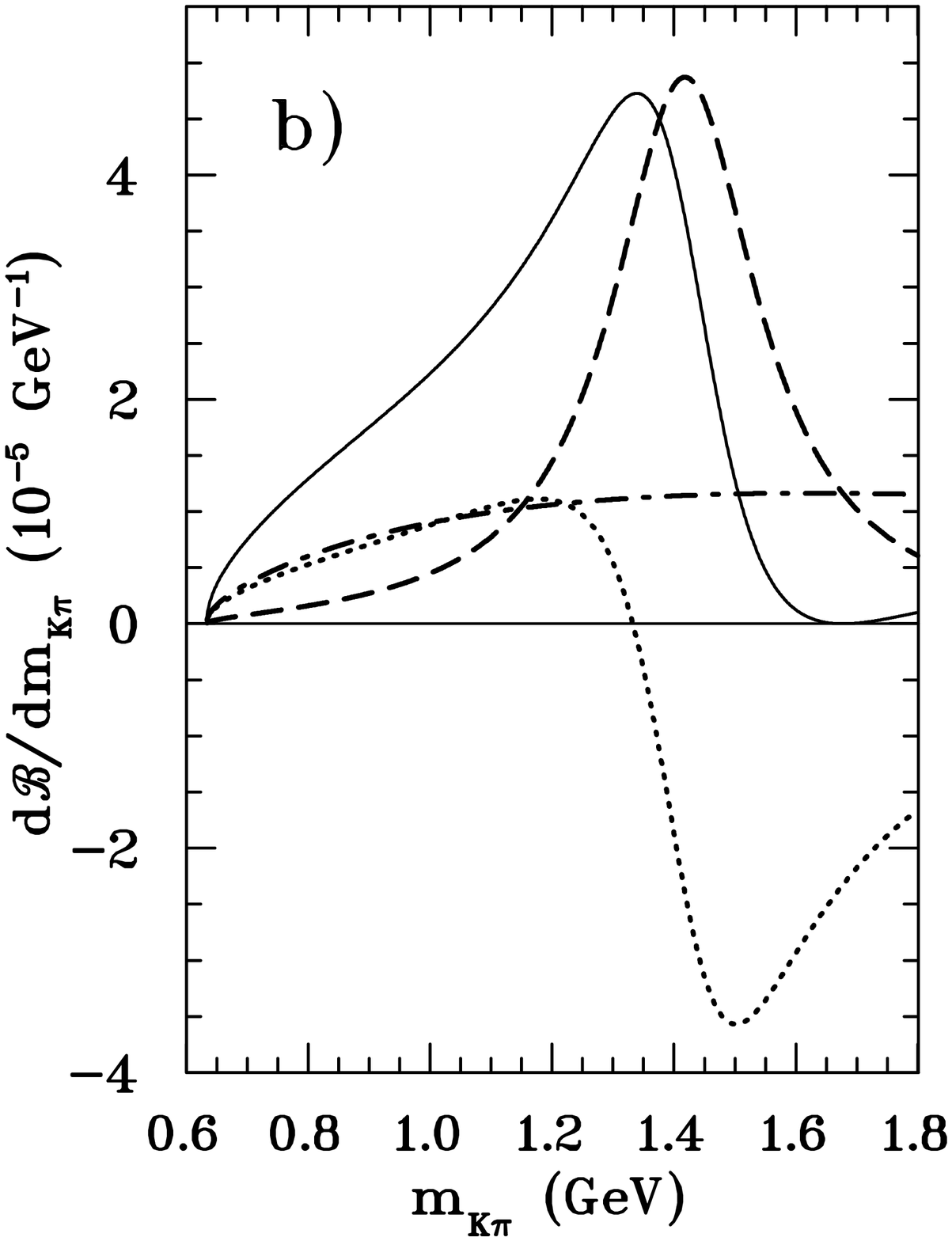}\\  
  \caption{Different components of the averaged $m_{K\pi}$ distributions of the $B^\pm \to (K^\pm\pi^\mp)_S
  \pi^\pm$ decays for: a) our model and b) BaBar's LASS parameterization~\cite{Aubert:2008bj}.
The dashed lines correspond to the resonant $K^*_0(1430)$ contributions, the dotted-dashed lines to the background, dotted lines to the interference and the solid lines to their sum.
} 
\label{Fig.components}
\end{center}
\end{figure}

In Fig. \ref{Fig.components} we show decomposition of the averaged $m_{K\pi}$ distributions 
of the charged B meson decays into a resonant, background and interference terms.
We compare these components calculated using our amplitudes and the BaBar's LASS
parameterization.
The background part in our approach is several times smaller than that effective range term
in experimental analyzes.
Worthy of note is the fact that interference parts have very different sizes and even 
different signs after integration over $m_{K\pi}$.



\section{Conclusions}

We have shortly presented main results of our analysis of the $K\pi$ interactions in 
the $B \to K\pi^+\pi^-$ decays with $K\pi$ pairs in the $S$- and $P$-waves.
In our approach sufficiently good description of experimental data can be achieved after introduction 
of four complex phenomenological coefficients into amplitudes.
These additional parameters are added to the QCD factorization coefficients and
can be interpreted as coming from long-distance charming penguin amplitudes, hard
spectator and weak annihilation contributions.

We have shown that the $K\pi$ $P$-wave amplitude is entirely dominated by 
$K^*(892)$ whereas the $S$-wave with the wide resonance $K^*_0(1430)$ and the very wide 
enhancement below 1 GeV (often interpreted as the $K^*_0(800)$ meson) is more complex. 
In order to correctly account for these resonances
we use scalar and vector form factors in the $S$- and $P$-waves
instead of the Breit-Wigner amplitudes, often employed in the isobar model.
This leads to the background term in the effective mass distributions 
which is several times smaller than the one obtained in each experimental analysis.

We therefore propose to use in the future
experimental analyzes the following parameterization of the $S$-wave 
$B \to (K\pi)_S\pi$ decay amplitude:
\begin{equation}
\mathcal{A}_S^- =  f_0^{K\pi}(m_{K\pi}^2)(c_0/m_{K\pi}^2 + c_1),
\nonumber
\end{equation}
where $f_0^{K\pi}$ is the scalar $K\pi$ form factor, while $c_0$ and $c_1$ are
complex numbers to be fitted from the data. 
Numerical values of the complex scalar form factor can be provided on request.

Let us emphasize at the end that the amplitudes derived in our analysis \cite{BFKLLM} go 
beyond the two-body approach applied to such decays as $B\to K^*\pi$.
We have proved that they correctly incorporate the two body final state interactions between 
$K$ and $\pi$ if the effective $K\pi$ mass in the three-body $B \to K \pi^+\pi^-$ decays is limited to about 2 GeV.

\end{document}